\documentclass[structabstract]{aa}
\usepackage{graphicx}
\usepackage[varg]{txfonts}
\usepackage{longtable}
\usepackage{lscape}
\usepackage[authoryear]{natbib}
\bibpunct{(}{)}{;}{a}{}{,} 

\begin{document}
\title{On the classification of flaring states of blazars}

\author{E.~Resconi\inst{1} \and D.~Franco\inst{2} \and A.~Gross\inst{1}\inst{3} \and L.~Costamante\inst{4}   
\and E.~Flaccomio\inst{5} }

\institute{Max-Planck-Institut f\"ur Kernphysik, Saupfercheckweg 1, 69117 Heidelberg, Germany \and
Dipartimento di Fisica, Universita degli Studi e INFN, 20133 Milano, Italy \and
University of Canterbury, Private Bag 4800, Christchurch New Zealand \and
Stanford Univeristy, W.W. Hansen Experimental Physics Laboratory and Kavli Institute for Particle Astrophysics and
Cosmology,Standford, CA 94305-4085, USA \and
INAF, Osservatorio Astronomico G.S. Vaiana, Piazza Parlamento I. 90134 Palermo, Italy 
}

\date{Received / Accepted}

\abstract {}
{The time evolution of the electromagnetic emission from blazars, in particular 
high frequency peaked sources (HBLs), displays irregular activity not yet understood.
In this work we report a methodology capable of  
characterizing the time behavior of these  variable objects.}
{The Maximum Likelihood Blocks (MLBs) is a model-independent estimator which 
sub-divides the light curve into time blocks, whose length and amplitude are compatible
with states of constant emission rate of the observed source. 
The MLBs  yields the statistical significance in the rate
variations and strongly suppresses the noise fluctuations in the light curves.}
{We apply the MLBs  for the first time on the long term X-ray light
curves  (RXTE/ASM) of Mkn~421,
Mkn~501, 1ES~1959+650 and 1ES~2155-304, which consist of more than 10 years of observational data
(1996-2007).  Using the MLBs interpretation of RXTE/ASM data, the integrated time flux distribution 
is determined for each single source considered. 
We identify in these distributions the characteristic level  as well as
the flaring states of the blazars
All the distributions show a significant component at negative flux values, most probably caused by  
an uncertainty in the background subtraction and by intrinsic fluctuations of RXTE/ASM. 
This effect interests in particular 
short time observations. In order to quantify the probability that the intrinsic fluctuations give rise to
a false identification of a flare,
we study a population of very faint sources and 
their integrated time flux distribution. 
We determine duty cycle or fraction of time a source spent in the flaring state of the source Mkn~421,
Mkn~501, 1ES~1959+650 and 1ES~2155-304. Moreover, we
study the random coincidences between flares and generic sporadic events 
such as high energy neutrinos or flares in other wavelengths. }{}

\keywords{X-ray: observations, galaxies: quasars}

\maketitle

\section{Introduction}
Blazars are defined as Active Galactic Nuclei (AGNs) dominated by a highly variable component of non-thermal 
radiation produced in relativistic jets pointed close to the line of sight 
\citep{Begelman, Urry-Padovani}. 
One of their main characteristics is the flux variability on different time-scales: 
from  fast flares lasting few minutes to  high states of several months.
Blazars  are considered to be sites of energetic particle production  and potential sources of
cosmic rays up to energies of at least  $~10^{19}$~eV. \\
The standard blazar spectral energy
distribution (SED) shows two prevalent components: a hump at low-energy which peaks in the
frequency range between infrared and X-ray bands, and a second hump  at higher energy, 
 proportionally shifted in the range from MeV up to TeV $\gamma$-rays.
Two potential scenarios, the so-called {\it leptonic} and {\it hadronic} ones, have been
proposed in order to model the SED.  
In  {\it leptonic models}  \citep[e.g][]{Jones, Ghisellini, Mast}, synchrotron emission  from relativistic electrons is
responsible for the first hump.  Electrons in the jet plasma up-scatter low-energy photons to
high energies  via Inverse Compton, producing the second hump. In this scheme, the 
same electron population produces both components. 
In {\it hadronic models}
\citep[e.g][]{Mannheim1,Mannheim2,felix1} protons are accelerated in the  jet together with electrons. 
The
synchrotron radiation produced by primary and proton-induced electrons 
contribute to the low-energy component. High-energy radiation originates
from photo-meson interactions and from proton and muon synchrotron radiation.  A
comprehensive description of a Monte Carlo simulation of a stationary synchrotron proton blazar
model,  including all relevant emission processes, can be found in \cite{Anita}. 
In hadronic models, $\gamma$-ray production by pion photo-production  results in  simultaneous
neutrino production. The decay of charged pions is the main neutrino production channel as
discussed in \citep{AnitaII}. \\
The detection of very high energy neutrinos coming from blazars
would be an unambiguous proof  of the existence of baryonic loaded outflows and would indicate
that blazars accelerate high energy cosmic rays.  Neutrino telescopes 
\citep[e.g.][]{icecube, icecuberes, baikal}
 until
now have not detected any extraterrestrial source of  neutrinos in the TeV-PeV energy region. 
As discussed in \cite{Rachen}, {\it \textquotedblleft the transience of energetic emission could improve the association of 
detected neutrinos with their
putative sources, because one could  use both arrival direction and arrival time information, allowing
statistically significant statements even for total fluxes below the background level"}.
This is true under the assumption that neutrino production in HBLs is subjected to the same
mechanisms at the base of the electromagnetic activity. Consequently, neutrino production and
electromagnetic activity should show the same time modulation.
The observation of time coincidences between electromagnetic flares and rare events, like neutrinos,
represents a natural test to the hadronic scenario. \\
The main requirement to this approach is a
clear definition and classification of the states of activity of the observed source.
In this paper we discuss 
 a procedure able to identify {\it characteristic}
and {\it flare} states in a light curve. The estimator that best fits our
requirements is the Maximum Likelihood Blocks (MLBs), since it is model-independent, it has been
designed to identify blocks of data with a constant rate in variable periods, and it provides a
statistical significance for each block. 
To test our approach, we perform a complete and detailed analysis on  RXTE/ASM X-ray light
curves. In particular, we analyze data from the  brightest High energy peaked BLLacs (HBLs) 
\citep{Giommi}: Mkn~421,
Mkn~501, 1ES~1959+650 and 1ES~2155-304.  
In the first part of this paper we describe the MLBs and how to  
separate {\it flares} from the {\it characteristic} level. 
Moreover, we introduce a definition for the duty cycle of the source. 
In the second part, we discuss the application of the method on RXTE/ASM data.

\section{Methods}
A variety of methods are used in astrophysics in order to assess the variability of a source and to qualify the
 character of the variability (periodic, correlated etc). 
It is not our intention here to review these methods.
Each method is designed for a specific purpose. 
Often, data are affected by large uncertainties or the data spacing is rather inhomogeneous.
The driving factors for the selection
of a method are the goals of the analysis and the quality of the data to be analyzed.
In our case we need a method that addresses the variability issue on light curves which 
are unevenly spaced and have short and long breaks, 
takes into consideration the statistical errors and possible unknown instrumental fluctuations on the measurements
and gives a representation of the light curve in term of periods in which the data points are compatible with a 
constant level. A method that could satisfy these requirements is the Maximum Likelihood Blocks.
The entire data analysis reported here is 
performed in ROOT \citep{root},
an object-oriented data analysis framework.
The only exception is for the Maximum Likelihood Blocks algorithm which is currently an IDL based program.

\subsection{Representation of the light curve: Maximum Likelihood
Blocks}

The methods used in the study of temporal variability depend strongly
on the nature of the available data and of the signal of interest. In
all cases, the most basic step is the classification of the
time-series as ``constant'' or ``variable''. Suitable and widely used
statistical tests include the Kolmogorov-Smirnov test for time-tagged
data (e.g. the arrival times of X- and $\gamma$-ray photons) and the
$\chi2$ test for binned data (e.g. binned photon arrival times or
optical magnitudes). The next step in the analysis of light curves is
the characterization of their ``shape''. We will  use a simple and
model-independent approach that aims at dividing the light curve into
time intervals in which the source emission is compatible with a
constant level. An algorithm based on Bayesian statistics that
performs such a segmentation for data of different natures was
presented by \cite[S98, ][]{scar98}. In its form for
time-tagged data, this algorithm was used for example to characterize
the X-ray light curves of young pre-main sequence stars observed by
the {\em Chandra} Orion Ultra-Deep Project \citep[COUP, ][]{get05} in
the Orion Nebula Cluster (ONC). A modified version of the S98
algorithm, based on a Maximum Likelihood rather than a Bayesian
approach, was recently employed in other studies of stellar X-ray
light curves \citep{wolk05,fava05,stel06}. We will refer to this
algorithm as the Maximum Likelihood Blocks (MLBs) and we introduce
here a variant that is suitable for the analysis of binned light
curves.\\
\noindent
Our algorithm is derived from the one presented by S98. Like
S98, we tackle the problem of finding the {\em best} piecewise
representation of a binned light curve in an iterative (and
approximate) way: we begin by testing the data against a constant-flux
model. If the model does not represent the data adequately we split
the light curve into two parts at the most likely {\em
\textquotedblleft change point"}. We then repeat these two steps
recursively on the resulting segments until all segments are
compatible with constant emission. The fundamental difference
with the algorithms presented by S98 lies in the statistics used to
test if a light curve is variable and to find the most likely change
point: rather than ``marginal likelihoods'' and ``Bayes factors''
(e.g. Eq.~7 and 48 in S98) we employ simple Likelihood functions, i.e.
the probability densities of obtaining the observed data set given a
parametric model. \\
\noindent
As mentioned above the algorithm was first applied to time-tagged
data. It is here adapted to the different statistical properties of
binned time series. Our lightcurves can be described as a series of
independent flux measurements, $r_i$, each normally distributed about
their mean values with standard deviations $\sigma_i$. The likelihood
of a parametrized model, $M$, of the lightcurve is maximized by
minimizing the $\chi2 = \sum (r_i - r_i[M])2/\sigma_i2$), where
$r_i[M]$ are the model-predicted fluxes.  In our case the model $M$ is
either the single-segment representation or one of the possible
two-segment representations of the light curve. We will refer to these
models, respectively, as  ``1'', specified by one parameter, the
constant flux level, and ``2(j)'' specified by three parameters: the
change point ``(j)'' (more specifically the index of the last point in
the first of the two segments) and the two flux levels. In this
notation our algorithm reduces to: $i$) splitting the light curve if
the minimum $\chi_12$ is such that the probability of obtaining a
larger value is lower than a user defined confidence threshold (e.g.
1\% or 0.1\%); $ii$) choosing the change point, $j_{cp}$, as the one
that minimizes $\chi_22$\footnote{The minimization with respect to
the flux levels is trivial and reduces to choosing the mean flux in the
given time interval.}.

\subsection{Interpretation of the light curve: flares versus characteristic level}
 
The goals of this analysis are the identification of  the various levels of activity of a source and the separation between
bursting events ({\it flares}) and steady state period(s) ({\it characteristic level(s)}). 
Sometimes periods of no variable activity are defined in the literature as {\it \textquotedblleft quiescent"}.
As discussed for example in \cite{wolk05},
the meaning of {\it quiescent} emission is ambiguous. An apparently {\it quiescent} level can be due to 
a superposition of numerous unresolved flares.  {\it Quiescent}, as defined as inactive, 
is therefore not appropriate to describe the level of activity in which the source spends most of the time.
We define the  {\it characteristic} level as $R_{char}$ and the spread around it  $\sigma_{char}$.
In order to determine the value of $R_{char}$ 
we construct the distribution of the amplitude $r_i$
and the duration of the single block $T_i$. We call this  
{\it integrated  time(T)-flux(r) distribution} based on the MLBs interpretation (B): $T_B(r)$.
This provides the distribution of the total amount of  time the source passes in a particular activity state.
The threshold above which a flux state 
is defined as {\it flare} is then defined as $R_{N\sigma}=(R_{char} +N\sigma_{char})$.
Depending on $N$, the probability that a selected flare state is   
 caused by a fluctuation of the characteristic level, by 
 an instrumental fluctuation or by a real enhancement of the photon emission from the source
  can be fully assessed.\\
On the base of this definition of {\it flares} we can determine
as well the frequency of 
flare states or {\it duty cycle $D_{N\sigma}$} as:

\begin{equation}
D_{N\sigma} = \frac{\int_{R_{N\sigma}}^{\infty} T_B(r)dr}{\int_{0}^{\infty} T_B(r)dr}
\end{equation}

In Sect. 4, the application of this method to RXTE/ASM data are reported.

\section{Data}
The All-Sky Monitor (ASM) on board of the Rossi X-ray Timing Explorer 
(RXTE) has been monitoring the X-ray sky routinely since March 1996. 
During each orbit up to 80\% of the sky is surveyed to a depth of 20-100 mcrab. 
A source is observed roughly 10 times a day.
A set of linear least-square fits over 90 seconds observation periods,
 by one of the three Scanning Shadow Cameras, yields 
the source intensities in four energy bands (1.5-3, 3-5, 5-12, and 1.5-12 keV). 
The intensities are usually given in units of the count rates expected if the sources were at the center 
of the field of view in one of the cameras. In 1.5-12 keV band, the Crab Nebula
 flux corresponds to about 75 ASM counts per second.  
A detailed description of  ASM can be found in \cite{ASM}. RXTE {\it standard data products} 
are collected directly from the HEASARC database.\\
We concentrate our study on RXTE/ASM data because
this provides the longest light curves in X-ray of Mkn~421,
Mkn~501, 1ES~1959+650 and 1ES~2155-304. 
However, for these kind of sources, the RXTE/ASM 
sensitivity is limited and data are affected by large errors.   
Moreover, the resolution and the background level of ASM observations depend on the 
Sun contamination or  back-scattered solar X-rays and on the detector stability along the 10
years of data taking, \cite{period_ASM}.

\section{Results}
\begin{figure*}
\centering
\includegraphics[width=17cm]{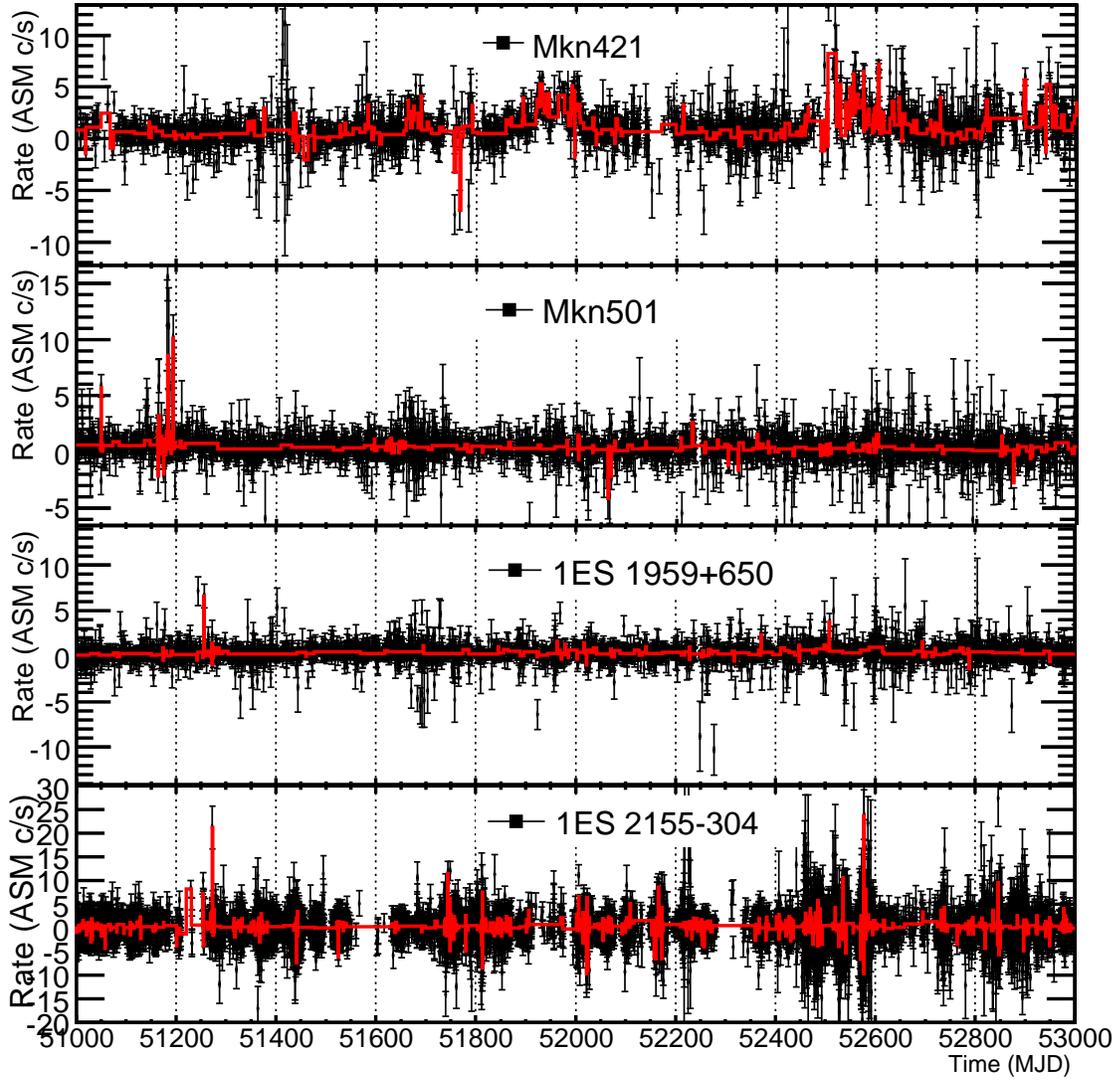}
\caption{\label{Figure_1}  Comparison between RXTE/ASM light curve (black points) and the MLBs 
(red line) for Mkn~421, Mkn~501, 1ES~1959+650, 1ES~2155-304. The period of time reported in the
figure is a sub-period respect the one analyzed; data considered in the paper have been collected 
for the period MJD 50200-53698 that corresponds to nearly 10 years. }
\end{figure*}

\begin{figure}
\resizebox{\hsize}{!}{\includegraphics{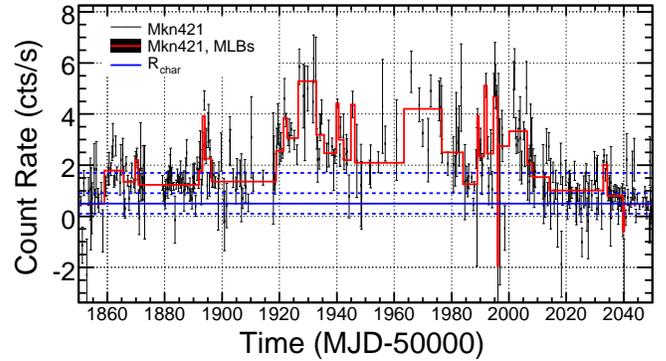}}
\caption{ RXTE-ASM dwell by dwell data for Mkn~421 in a period of time 
where the source was particularly active. Results of
MLBs are reported in red. The blue continue line corresponds to the
characteristic level $R_{char}$ of the source.
 The blue dashed lines represent the $1\sigma_{char}$ and the $3\sigma_{char}$ spread
 around $R_{char}$. States above $R_{1\sigma}$ are active states and states above 
 $R_{3\sigma}$ can be considered very high states.}
\label{Figure_2}
\end{figure}

The results of the application of the MLBs to the RXTE/ASM data for the four HBLs considered are
 reported in    Fig.~\ref{Figure_1} and in Fig.~\ref{Figure_2}. Each {\it change point} identified by the algorithm has a
statistical significance of at least $3 \sigma$. 
This operation is still not enough in order to characterize the 
behavior of a source.
We haven't yet quantified the
characteristic noise of RXTE/ASM, then we can not distinguish if 
the change points identified by the MLBs are  
of instrumental nature or of a physics nature.  
In order to distinguish between these two scenarios and study
the behavior of sources, 
we construct the 
 {\it integrated time(T)-flux(r) distribution} $T_B(r)$, as discussed in Sect. 2.2.
We first construct the $T_B(r)$ for a set of very faint sources in order to study
the intrinsic fluctuations of the instrument 
and finally we determine the $T_B(r)$ for Mkn~421, Mkn~501, 1ES~1959+650
and 1ES~2155-304. On the base of these distributions we identify the states that can be
considered {\it flares} with  good confidence.

\subsection{RXTE/ASM intrinsic fluctuations}
\begin{figure*}
\centering
\includegraphics[width=17cm]{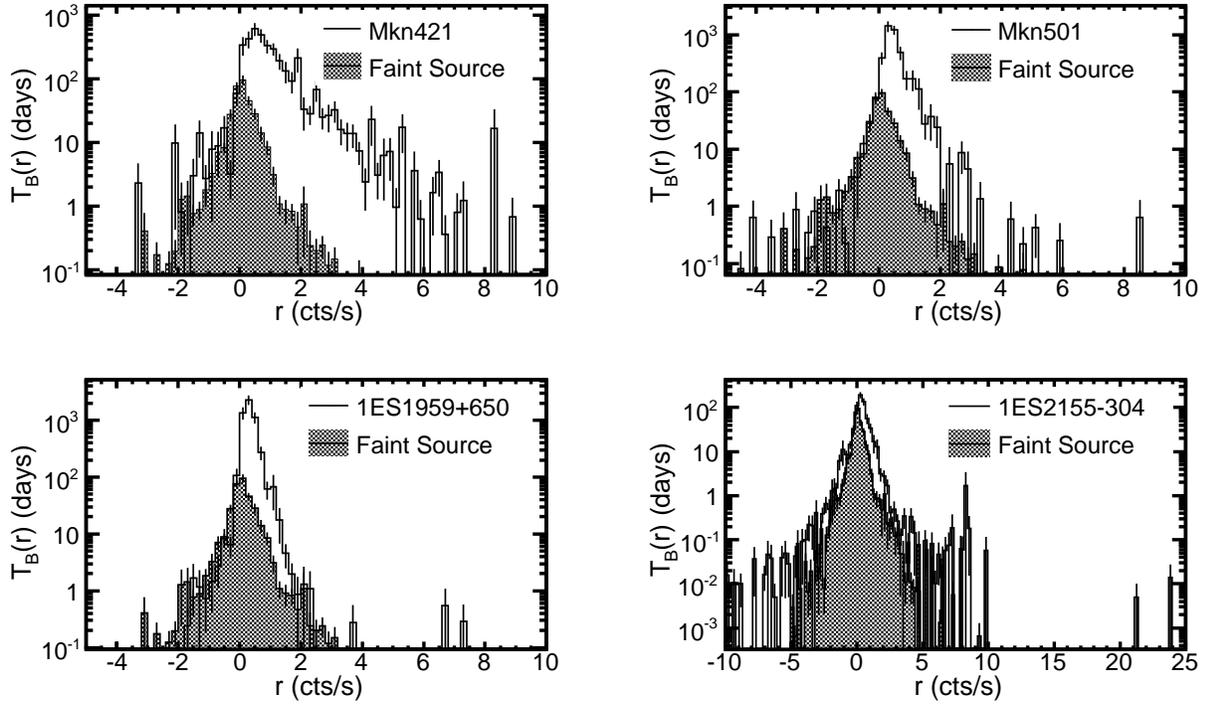}
\caption{\label{Figure_3} Integrated time flux distribution for Mkn~421 (top-left), Mkn~501 (top-right), 1ES~1959+650 (bottom-left) and
1ES~2155-304 (bottom-right).The distributions are compared with the integrated time flux distribution of a very faint source (dark
background).}
\end{figure*}

In Fig.~\ref{Figure_1}, we notice that the MLBs identify not only significant {\it change points} 
at positive amplitudes but also at negative ones. 
These negative fluctuations can be caused by uncertainties in the background subtraction 
and by intrinsic fluctuations of RXTE/ASM.
In order to characterize such a component and its effect on the definition of flares,
 we have analyzed RXTE/ASM light curves for a set of very faint sources, since
these are expected to spend most of their time at a flux level 
well below the ASM sensitivity. 
The sources considered, reported in Tab.~\ref{tab:ASM bg}, 
have a low X-ray monochromatic average emission (less then 0.6 $\mu$Jy at 1keV), 
are randomly distributed in the sky and are at 
various redshifts.
The flux distributions of the faint sources are all normal distributions, as expected for a random
instrumental noise. On average, the normal distributions peak at rate $r \approx 0.1$ ASM c/s and have 
a standard deviation of $r \approx 0.3$ ASM c/s. Since the studied faint sources show similar flux
distributions, we will use in the next just one of them for comparison; the 
source used is PKS~0118-272 and represent the average faint source in our sample.
In Fig.~\ref{Figure_3}, the flux distribution of PKS~0118-272
is compared with 
the flux distribution of the HBLs considered in this work.
The distributions are normalized using the areas under the negative flux tails.
In this way, we estimate the fraction of the HBLs flux distributions caused by 
the intrinsic fluctuations of RXTE/ASM ($S_{N\sigma}$).

\subsection{RXTE/ASM flare states}
The $T_B(r)$ for the Mkn~421, Mkn~501, 1ES~1959+650
and 1ES~2155-304 are reported in Fig.~\ref{Figure_3}. 
All the distributions differ significantly from a normal distribution 
indicating that RXTE/ASM is indeed sensitive to high activity states of the sources considered. 
For all the HBLs considered  we observe that the flux distribution 
shows a peak above the pure background distribution. 
We define the central peak value (that corresponds also to the {\it mode} of the flux distribution) 
and its standard deviation as $R_{char}$ and $\sigma_{char}$. 
More sophisticated fitting procedures have been applied but, given the quality of the data of RXTE/ASM, 
they did not provide a more precise estimation of $R_{char}$ and    
 $\sigma_{char}$.  We observe that $R_{char}$ is next to the detection threshold of RXTE/ASM.
  As discussed in Sect. 2.2,  $R_{N\sigma}=(R_{char} +N\sigma_{char})$
 is the threshold above which a flux state 
is considered a {\it flare}.\\
Using this definition of {\it flares} we determine
the frequency of flare states or {\it duty cycle $D_{N\sigma}$} as described in Sect. 2.2.
In Tab.~\ref{tab:ASM flares} we report $D_{N\sigma}$ for the HBLs considered and the cases of N=1 and N=3. 
For the specific case of RXTE/ASM, we calculate as well
the intrinsic fluctuation $S_{N\sigma}$ that affects the duty cycle as:

\begin{equation}
S_{N\sigma} = \frac{\int_{R_{N\sigma}}^{\infty} T_{B_{sys}}(r)dr}{\int_{R_{N\sigma}}^{\infty} T_B(r)dr}
\end{equation}
 
\noindent
where $T_{B_{sys}}(r)$ is the  flux distribution of a faint source (in this work PKS~0118-272).
Results are reported in Tab.~\ref{tab:ASM flares}.

\subsection{Example of application: correlation study between electromagnetic flares and neutrinos }
\begin{figure}
\resizebox{\hsize}{!}{\includegraphics{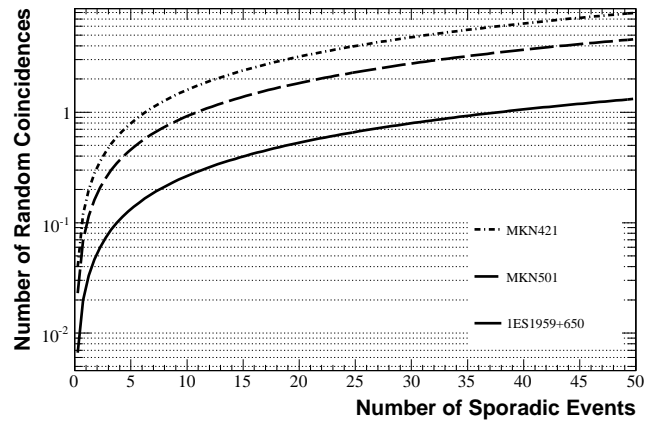}}
\caption{Distributions of random coincidences  
 between Mkn~421, Mkn~501, 1ES~1959+650 flare states above 3$\sigma_{char}$ and N
neutrinos or more general sporadic events like rare flares in other wavelength. }
\label{Figure_5}  
\end{figure}

As anticipated in the introduction,
the study of the physics of HBLs develops through different approaches. 
One of the most frequently used  
sees the study of flare
correlation  among different wavelengths,  for example X-ray and TeV-$\gamma$ rays \cite{LM}. 
A study of the correlation among different messengers such as photons and neutrinos is 
a more recent interest \cite{ic} and has been the motivation of this work. 
The significance of such correlations can be assessed only
when the frequency of the electromagnetic flare states is determined, for example
following the procedure described in this paper. 
In order to illustrate such a case, we study the distribution 
of coincidences between RXTE/ASM flare states and 
a set of  N neutrinos or equivalent sporadic events. 
The flares are selected following the procedure described above 
for $3\sigma_{char}$  and
the N neutrino events are uniformly distributed in the entire time period 
considered in this paper, 
approximately 10 years.
The distributions of coincidences between the
 RXTE/ASM flare periods for Mkn~421, Mkn~501 and 1ES~1959+650 
and the neutrino events are reported in Fig.~\ref{Figure_5}. 
Depending on the number N of sporadic events
we are able to determine the number of random coincidences 
multiplying the probability of a flare, or duty cycle of the source, and N. 
Once the random coincidence among flares and neutrinos is determined, then
the statistical meaning of experimentally observed coincidences between flares and neutrinos  
can be determined. This eventually can hint at the association of detected neutrinos with 
an  astronomical source even if N is below or at the level of the expected background.

\section{Conclusions} 
The X-ray time behavior of Mkn~421, 1ES~1959\-+650, Mkn~501 and 1ES~2155\--304 has been
 characterized using approximately 10 years of data from RXTE/ASM. 
The characteristic level and flaring states have been defined and values are reported in
Tab.~\ref{tab:ASM flares}.\\
Mkn~421 is the source that flares most often amongst those studied:
 for $\sim40\%$ of the RXTE/ASM observations
the source was in an active state (above $R_{1\sigma}$) 
 and for $\sim 18\%$ of these it was in a very high state (above $R_{3\sigma}$).
The probability that a flare state is caused by a fluctuation of the instrumental noise 
is marginal. This confirms the well known fact that Mkn~421 is an extremely variable HBL and 
quantifies for the first
time its duty cycle even if this is only valid for RXTE/ASM.
Mkn~501 flares often at low rates  ($\sim26 \%$) and less often at high rates ($\sim10 \%$).
Also in this case, the intrinsic fluctuations  do not significantly affect the flare states. 
1ES~1959+650 flares more rarely in particular at high rates (only 2.6\%). The systematic component
affects one tenth of flares.
In the case of 1ES~2155-304 nearly one fourth of flare is caused by an intrinsic fluctuation of RXTE/ASM. 
This simply means that this source
is at the threshold limit for RXTE/ASM.  \\
The study of the significance of correlations 
of flares among different wavebands and different messengers is foreseen for a future work.

\begin{table*}
\caption{HBLs and X-ray faint sources considered in this work. The first 
column contains the X-ray monochromatic emission, the second and the third contain the 
mean and standard deviation of the Gaussian that fit at the best the flux distribution. 
Last column contains the red-shift. The mean value of
the RXTE/ASM count rate is reported for the faint sources only since the HBLs
 are too variable for a definition
of a mean flux value. For comparison, the resulting $R_{char}$ and $\sigma_{char}$ 
are reported in Tab.~\ref{tab:ASM flares}.}
\label{tab:ASM bg}
\centering
\begin{tabular}{l|cccc}
\hline\hline
Source Name	   & 	X-ray at 1keV ($\mu$Jy)  	
& Mean (ASM c/s)   &   Sigma (ASM c/s) & z\\
\hline
Mkn~421  &    10.1 & & & 0.031\\
Mkn~501  &	   9.5 & & & 0.034\\
1ES~1959+650  &	   9.4 & & & 0.048\\
1ES~2155-304  &  15.2 & & &0.117\\
\hline
PKS~0118-272 	& 0.22 & 0.03 & 0.25 & 0.557\\
1ES~0235+164 	& 0.28 & 0.09 & 0.24 & 0.94\\
S5~0454+844 	& 0.03 & 0.07 & 0.21 & 0.112\\
PKS~0735+178	& 0.20 & 0.11 & 0.31 & 0.424\\
PKS~0829+046	& 0.08  & 0.07  & 0.25 & 0.174\\
S4~0954+65	& 0.56 & 0.09 & 0.21 & 0.368\\
B2~1147+245 	& 0.04 & 0.10 & 0.24 & $\sim$0.2\\
\hline
\end{tabular}
\end{table*}

\begin{table*}
\caption{For each HBLs considered we give the characteristic level $R_{char}$ and its 
standard deviation $\sigma_{char}$. The duty cycle for active state
$D_{1\sigma}$ and for very high states $D_{3\sigma}$ is also reported together with the 
relative intrinsic fluctuations still present above the $R_{char}+1\sigma$ or $R_{char}+3\sigma$ 
threshold due to RXTE/ASM
($S_{1\sigma}$, $S_{3\sigma}$).}
\label{tab:ASM flares}
\centering
\begin{tabular}{l|cccccc}
\hline\hline
Source Name & 	$R_{char}$ (ASM c/s)	
& $\sigma_{char}$ (ASM c/s)  & $D_{1\sigma}$ (\%) & $S_{1\sigma}$ (\%)
& $D_{3\sigma}$ (\%) & $S_{3\sigma}$ (\%) \\
\hline
Mkn~421  &    0.5 & 0.4 & 40.3   &   1.0 & 18.1  & 0.5 \\
Mkn~501  &    0.4 & 0.2 & 25.8  & 3.0 & 9.7  & 2.3 \\
1ES~1959+650  &	0.3 & 0.2 & 14.2  & 7.1 & 2.6  & 9.5\\
1ES~2155-304  & 0.2 & 0.4 & 27.0  & 22.1 & 4.9  & 18.8\\
\hline
\end{tabular}
\end{table*}

\begin{acknowledgements}
E.R. and A.G. are funded by the Deutsche Forschungsgemeinschaft (DFG) through an 
Emmy Noether grant (RE 2262/2-1). This research has made use of data
obtained through the High Energy Astrophysics Science
Archive Center Online Service, provided by the NASA/
Goddard Space Flight Center. We acknowledge the RXTE/ASM and RXTE/PCA team
 for the X-ray data. Thanks to A. Taylor, S. Movit and S. Odrowski 
 for their proofreading and useful comments.
\end{acknowledgements}

\bibliographystyle{aa}

\end{document}